\documentclass[12pt,a4paper]{article}
\usepackage{amsmath}
\usepackage{amssymb}
\usepackage{color}
\usepackage{graphicx}
\usepackage{float}
\usepackage{cite}
\usepackage{dcolumn} 
\newcolumntype{.}{D{.}{.}{1}}
\usepackage[ruled, vlined, linesnumbered]{algorithm2e}
\usepackage{algpseudocode}              
\usepackage{dsfont}
\usepackage{mathrsfs}
\usepackage{booktabs,threeparttable}
\usepackage{mhchem}
\usepackage{braket}

\usepackage{paralist}
\usepackage{xcolor}
\usepackage[normalem]{ulem}

\newcommand{\ckdel}[1]{\textit{\color{black!40!green!100}{\xout{#1}}}}
\renewcommand{\ckdel}[1]{}

\begin{document}
\begin{center}

\vspace*{1cm}

{\LARGE\bf
Adaptive Density-Guided Approach to
Double Incremental Potential Energy Surface Construction
}

\vspace{2cm}

{\large Denis G. Artiukhin$^{\dagger}$\footnote{Email: artiukhin@chem.au.dk}, 
Emil Lund Klinting$^{\dagger}$,
Carolin K\"onig$^{\parallel}\footnote{Email: koenig@pctc.uni-kiel.de}$,
\\ and Ove Christiansen$^{\dagger}$\footnote{Email: ove@chem.au.dk} \\[2ex]
}

\vspace{1.5cm}

$^{\dagger}$
Department of Chemistry, Aarhus Universitet, DK-8000 Aarhus, Denmark \\[1ex]

\vspace{0.5cm}

$^{\parallel}$
Institute of Physical Chemistry, Christian-Albrechts-University Kiel, \\ Max-Eyth-Stra{\ss}e 1, D-24118 Kiel, Germany\\[1ex]

\end{center}

\vfill

\begin{tabbing}
Date:   \quad\= \today \\
\end{tabbing}.

\newpage

\begin{abstract}

We present a combination of the recently developed double incremental  
expansion of potential energy surfaces with the well-established adaptive density-guided 
approach to grid construction. This unique methodology is based on the use of an incremental 
expansion for potential energy surfaces, known as $n$-mode expansion, 
an incremental many-body representation of the electronic energy, 
and an efficient vibrational density-guided approach to 
automated determination of grid dimensions and granularity. 
The reliability of the method is validated calculating potential energy surfaces and obtaining fundamental
excitation energies for three moderate-size chain-like molecular systems. 
The results are compared to other approaches, which utilize 
static grid construction for supersystem and fragmentation calculation setups.
The use of our methodology leads to considerable computational savings for potential energy surface construction
and a major reduction in the number of required single point calculations
can be achieved, while maintaining 
a high level of accuracy in the resulting potential energy surfaces. Additional investigations indicate that 
our method can be applied to covalently bound 
and strongly interacting
molecular systems, 
even though these cases are known as being very unfavorable for fragmentation schemes. 
We therefore conclude that the presented methodology is a robust and flexible approach 
to potential energy surface construction, which introduces considerable computational savings 
without compromising the accuracy of vibrational spectra calculations.

\end{abstract}

\newpage
\clearpage

\section{Introduction \label{sec:intro}}

Vibrational spectroscopies are valuable and well-established techniques in studies of molecular 
reactivity and dynamics. These methods can provide vast information on different structural aspects 
and associated properties. The information, however, is not direct, but rather extracted from
measured vibrational frequencies and intensities. This raises an issue of experimental signal assignment,
which can be a non-trivial task and typically requires theoretical assistance. 
Many important molecular properties depend on quantum-mechanical interactions 
 on the atomic level. 
In these cases, a fully quantum chemical description is desirable, but is often extremely complicated 
due to a very high computational cost. The task 
becomes even more involved, if the target properties have to be calculated for a number of molecular
geometries as, for example, in calculations of potential energy surfaces (PESs). The PES, however, is
the cornerstone for providing detailed information on vibrational motion, reactivity, and photochemical
properties. 
The very high computational cost of quantum chemical PES calculations is caused by a strongly unfavorable
scaling with respect to the system size. This is due to i) scaling of single point (SP) electronic structure
calculations and ii) an increasing dimensionality of the PES and, as the result, a growing number of SPs
to be computed. Many fundamentally different approximations have been developed to address the
first issue leading to so-called linear-scaling (with the system size) generation of 
computational methods~\cite{galli1996,goedecker1999,goedecker2003,zalesny2011}.
The second issue is often alleviated with well-known direct dynamics approaches, 
which calculate SPs on the fly, or
by using adaptive schemes for choosing grid points (for example, see Refs.~\cite{sparta2009,richter2012,strobusch2014}).
However, these techniques alone are far
from enough to achieve a linear-scaling behavior in the overall PES construction.

The PES is an $M$-dimensional object with the number of vibrational modes 
$M = 3N_{\mathrm{nuc}} - 6$ and $N_{\mathrm{nuc}}$ being the number of nuclei and scales linearly with $N_{\mathrm{nuc}}$.
The computational cost of PES construction, however, 
scales exponentially because of the necessity to calculate couplings between all vibrational modes. 
As the result, electronic structure computations of fully-coupled PESs are feasible 
for only a few atoms. It is, therefore, a common  
strategy to restrict these mode--mode couplings using the $n$-mode 
expansion~\cite{jung1996,carter1997,bowman2003,rauhut2004,kongsted2006}, which is also known 
as High Dimensional Model Representation (HDMR)~\cite{rabitz1999} or cluster expansion~\cite{meyer2012}.
The use of this approach leads to a polynomial scaling in the PES construction and 
 allows for all-mode anharmonic PES calculations of moderate-sized molecules 
 (about 15--20 atoms). 
A further reduction of the computational
cost is possible by combining the $n$-mode expansion with screening 
techniques~\cite{benoit2004,rauhut2004,benoit2006,pele2008,benoit2008,seidler2009,cheng2014}, 
by calculating higher order couplings in an approximate 
manner~\cite{rauhut2004,yagi2007,rauhut2008,Rauhut2009,sparta2009_2,sparta2010,meier2013,schmitz2019}
or using a reduced vibrational space for larger systems~\cite{benoit2004,mackeprang2015,yagi2019}.
Because PESs are usually constructed from a grid of SPs, another strategy to reduce the computational cost is to
ensure that the grid of SPs is   
optimal, i.e., that they consist of as few points as 
possible and correctly describe the shape of the PES. 
The adaptive density-guided approach (ADGA)~\cite{sparta2009} 
fulfills these purposes and allows for routine and automatic 
grid construction.
It does not require any prior knowledge of the molecular system from the user and, therefore, has a 
number of advantages over the often used static grid approach.
The ADGA has proven to be a valuable tool for PES 
construction and received a number of extensions such as
the posibility to use arbitrary fit-basis functions for the analytical representation of the PES~\cite{klinting2018} and
more sophisticated algorithms for grid boundary extension~\cite{klinting2018}.
It was also applied in conjunction with multiresolution procedures of PES construction~\cite{sparta_mult2009},
derivative information~\cite{sparta2010},
and
Gaussian process regression (GPR) algorithms~\cite{schmitz2019}.

The recently reported double incremental expansion~\cite{koenig2016} exploits ideas of restricting
direct mode--mode couplings~\cite{jung1996,carter1997,bowman2003,rauhut2004,kongsted2006} together 
with fragmentation of a molecule into subsystems. Therefore, it
addresses both above-mentioned scaling problems and significantly reduces the computational cost. 
This methodology
offers a possibility for robust and accurate calculations of the PES for larger molecular systems, when 
being combined with localized vibrational coordinates. 
Ultimately linear-scaling in the computational cost of constructing the full PES 
 is achieved, when introducing an additional approximate transformation step effecting 
a limited part of the high-dimensional PES.
However, the double incremental expansion has a large drawback, as it was initially developed to be used 
with the static grid approach, where equidistant grids of SPs need to be pre-defined by the 
user. Its combination with the ADGA allows for a further significant 
reduction in the overall computational cost
and, most importantly, for a full automatization of PES calculations.
In this work, we aim to fill this gap and 
present a joined methodology, which incorporates the strengths of both computational approaches.
Additionally, we further investigate the double incremental expansion by studying the convergence for 
high orders of truncation and show that it can be used in cases where fragmentation schemes are usually quite unfavorable.

We briefly introduce the single and double incremental expansions in Sec.~\ref{sec:theory}, as well as 
the ADGA. We provide computational details in Sec.~\ref{sec:comp_details} for the results given in Sec.~\ref{sec:results},
where we discuss our novel methodology in comparison to other approaches and in application to a chain-like molecular system. 
A conclusion to the results and associated discussions is given in Sec.~\ref{sec:concl}.

\section{Theory \label{sec:theory}}

First, we briefly sketch the incremental $n$-mode and many-body expansions 
in Secs.~\ref{sec:n_mode} and \ref{sec:many_body}, respectively.
Their combination resulting in a double incremental expansion
is described in Sec.~\ref{sec:die}, where
additionally, the role of semi-local coordinates and the neighbor coupling approximation is outlined.
For more rigorous derivations and extensive discussions on these topics, 
the interested reader is referred to the original works on double incremental expansions~\cite{koenig2016} 
and flexible adaptation of local coordinates of nuclei (FALCON)~\cite{koenig2016_falcon}.
Finally, a description of the ADGA using the double incremental expansion in PES construction
is given in Sec.~\ref{sec:adga}.

\subsection{$n$-Mode Expansion \label{sec:n_mode} }

The idea of the $n$-mode expansion~\cite{jung1996,carter1997,bowman2003,rauhut2004,kongsted2006} 
lies in representing the total PES $V(\mathbf{q})$, which depends on a set of $M = 3N_{\mathrm{nuc}}-6(5)$ vibrational coordinates
$\mathbf{q} = \{q_1, q_2, \dots, q_M\}$, in a series such that 
\begin{multline}\label{eq:n_mode_exp_long}
    V(\mathbf{q}) = V^{0} + \sum_{i=1}^{M} [ V^{m_i} - V^{0}  ] \\ 
    + \sum_{j>i=1}^{M} \left\{  [V^{m_i m_j} - V^{0} ] -[V^{m_i} - V^{0} ] - [ V^{m_j} -  V^{0} ]   \right\} + \cdots,
\end{multline}
where lower-dimension cut-potentials are, e.g., given by 
\begin{align}
    V^{0} &\equiv V(0,\dots,0),  \label{eq:pot_cut_func_1}  \\
    V^{m_i} &\equiv V(0,\dots,0, q_{m_i} ,0,\dots,0),    \label{eq:pot_cut_func_2}   \\
    V^{m_i m_j} &\equiv V(0,\dots,0,q_{m_i},0,\dots,0, q_{m_j},0,\dots,0)   \label{eq:pot_cut_func_3} 
\end{align}
and depend on zero, one, and two coordinates, respectively.  
The constant $V^{0}$ is a reference point, which is usually chosen to be the absolute minima of the PES, but is not required to be so. 
Note that the equality in Eq.~(\ref{eq:n_mode_exp_long}) holds only if the order of expansion $n$ is equal 
to the number of vibrational modes $M$, whereas
truncation of this expression at lower orders leads to an approximate treatment of $V(\mathbf{q})$. 
In the latter case, only mode couplings up to a particular order are allowed, 
i.e., only interactions between pairs of modes or triples of modes, etc.
For example, truncation of the $n$-mode PES expansion at the first order leads to the expression
\begin{equation}\label{eq:n_mode_exp_trunc_1M}
    V(\mathbf{q}) \approx V^{(1)} (\mathbf{q}) =  V^{0} + \sum_{i=1}^{M} [ V^{m_i} - V^{0}  ],
\end{equation}
where $V^{(1)}(\mathbf{q})$ is an approximate 1M PES, which depends on the same number of vibrational coordinates $M$, 
as the exact PES $V(\mathbf{q})$. More generally, truncation at order $n$ would lead to an $n$M PES denoted as $V^{(n)}(\mathbf{q})$ and would provide 
an exact description of the PES if the condition $n=M$ is satisfied.

A more concise form of Eq.~(\ref{eq:n_mode_exp_long}) can be derived by introducing bar-potentials,
\begin{align}
    \bar{V}^{0} &\equiv V^{0}, \label{eq:pot_bar_func_1} \\
    \bar{V}^{m_i} &\equiv V^{m_i} - \bar{V}^{0},  \label{eq:pot_bar_func_2}  \\
    \bar{V}^{m_i m_j} &\equiv V^{m_i m_j} - \bar{V}^{m_i} - \bar{V}^{m_j} + \bar{V}^{0},  \label{eq:pot_bar_func_3}
\end{align}
which are generated from the cut-potentials of 
Eqs.~(\ref{eq:pot_cut_func_1})--(\ref{eq:pot_cut_func_3}) by subtraction of all lower-order bar potentials.
The expression for the $n$-mode expanded PES thereby reads
\begin{equation} \label{eq:n_mode_exp_short}
    V(\mathbf{q}) = \bar{V}^{0} + \sum_{i=1}^{M} \bar{V}^{m_i} + \sum_{j>i=1}^{M} \bar{V}^{m_i m_j} + \cdots.
\end{equation}
Following the description in Ref.~\cite{koenig2016}, we introduce  
mode combinations (MCs) $\mathbf{m}_k$, as sets containing $k$ coordinate indices.   
For example, a general MC of order two can be written as $\mathbf{m}_2 = \{m_i, m_j\}$. Using this set, the corresponding cut-potential
$V^{m_i m_j}$ from Eq.~(\ref{eq:pot_cut_func_3}) can be abbreviated as $V^{\mathbf{m}_2}$.
Combining all MCs from an $n$-mode expansion into a larger set, we define the  
mode combination range (MCR). Thus, the MCR for a PES depending on two coordinates $V(q_1, q_2)$ and expanded to the second order would be equal 
to $\{  \{\}, \{1\}, \{2\}, \{1,2\}   \}$, where
\{\} denotes an empty set. 
It is required that the MCR is closed on forming subsets.
This condition is satisfied if all subsets of each $\mathbf{m}_k \in \mathrm{MCR}$ are also included in the MCR. 
It is thereby possible to introduce a general expression for the bar-potentials of Eqs~(\ref{eq:pot_bar_func_1})--(\ref{eq:pot_bar_func_3}) as
\begin{equation} \label{eq:bar_funcs_general}
    \bar{V}^{\mathbf{m}_k} = V^{\mathbf{m}_k} - \sum_{ \substack{
        \mathrm{\bf m}_{k^{\prime}} \subset  \mathrm{\bf m}_k   \\
        \mathrm{\bf m}_{k^{\prime}} \in \mathrm{MCR}  }}  \bar{V}^{\mathbf{m}_{k^{\prime}}}, 
\end{equation}
where the sum runs over all subsets of $\mathbf{m}_k$, which are included into an MCR that is closed on forming subsets.   
Finally, the $n$-mode expansion of Eq.~(\ref{eq:n_mode_exp_short}) can be written in a shorter and more practical form,
\begin{equation}\label{eq:n_mode_exp_final}
   V(\mathbf{q}) = \sum_{ \mathbf{m}_k \in \mathrm{MCR} }   \bar{V}^{ \mathbf{m}_k  }     
        = \sum_{ \mathbf{m}_k \in \mathrm{MCR}  } \sum_{ \substack{
        \mathrm{\bf m}_{k^{\prime}} \subseteq  \mathrm{\bf m}_k   \\
        \mathrm{\bf m}_{k^{\prime}} \in \mathrm{MCR}  }}  (-1)^{k-k^{\prime}}  V^{ \mathbf{m}_{k^{\prime}}   }.
\end{equation}
This expression is exact, if $k$ runs over all values up to the number of vibrational modes $M$, whereas an approximate 
$n$M PES $V^{(n)}(\mathbf{q})$ is constructed if the series is truncated at the order $n$. In the latter case, $k$ takes on all values up to $n$. 
Note that this final expression, as seen in Eq.~(\ref{eq:n_mode_exp_final}), 
is simply a reformulation of the well-known $n$-mode expansion in terms of MCs. 
However, it provides the necessary flexibility for defining double incremental expanded PESs, as will be shown in Sec.~\ref{sec:die}.

\subsection{Many-Body Expansion\label{sec:many_body}}

Similar to the $n$-mode expansion, as described in Sec.~\ref{sec:n_mode}, we can derive a many-body expansion for the total energy $E$, understood as
an eigenvalue of the electronic Hamiltonian for
a molecular system. To this end, we consider our molecule as being composed of $N_{\mathrm{frag}}$ fragments. The molecular structure of fragment $i$ is 
described by a composite variable $z_i$, which represents the complete nuclear conformation of the fragment. Hence, the total energy $E$ can be written as a function of these variables such that
$E = E(z_1, z_2, \dots, z_{N_{\mathrm{frag}}})$. 
Analogous to Eq.~(\ref{eq:n_mode_exp_long}), the many-body expansion of the total energy can be written as 
\begin{multline}\label{eq:many_body_exp_long}
    E(\mathbf{z}) = E_{0} + \sum_{i=1}^{N_{\mathrm{frag}}} [ E_{f_i} - E_{0}  ] \\ 
    + \sum_{j>i=1}^{N_{\mathrm{frag}}} \left\{  [E_{f_i f_j} - E_{0} ] -[E_{f_i} - E_{0} ] - [ E_{f_j} -  E_{0} ]   \right\} + \cdots,
\end{multline}
where cut-functions are defined by the expressions,
\begin{align}
   E_{0} &\equiv E(0,\dots,0),   \\
   E_{f_i} &\equiv E(0,\dots,0, z_{f_i} ,0,\dots,0),   \\
   E_{f_i f_j} &\equiv E(0,\dots,0,z_{f_i},0,\dots,0, z_{f_j},0,\dots,0).
\end{align}
Here, $E(z_1, z_2, \dots, z_{N_{\mathrm{frag}}})$ should be understood as the total energy of the entire molecular system, whereas 
cut-functions $E_{f_i}$ and $E_{f_i f_j}$ are total energies of a single fragment $i$ and of two collective fragments $i$ 
and $j$, respectively.  
We set $E_{0}$ to be equal to zero similar to the case of $V^0$.
The equality in Eq.~(\ref{eq:many_body_exp_long}) holds only if the order of expansion $l$ is equal to the number of fragments $N_{\mathrm{frag}}$. 
Similar to the case of the $n$-mode expansion for PESs, we define an approximate energy $E^{(l)} (\mathbf{z})$, which result from 
the truncation of Eq~(\ref{eq:many_body_exp_long}) at order $l$ and define this level of approximation as $l$F.
By defining fragment combinations (FCs) $\mathbf{f}_j$ as sets containing indices $f_i$ and introducing 
fragment combination ranges (FCR) for FCs explicitly included in a many-body expansion,
we can recast Eq.~(\ref{eq:many_body_exp_long}) as 
\begin{equation} \label{eq:many_body_exp_final}
   E(\mathbf{z}) = \sum_{ \mathbf{f}_j \in \mathrm{FCR} }   \underline{E}_{ \mathbf{f}_j  }     \\
        = \sum_{ \mathbf{f}_j \in \mathrm{FCR}  } \sum_{ \substack{
        \mathrm{\bf f}_{j^{\prime}} \subseteq  \mathrm{\bf f}_j   \\
        \mathrm{\bf f}_{j^{\prime}} \in \mathrm{FCR}  }} (-1)^{j-j^{\prime}}   E_{ \mathbf{f}_{j^{\prime}}   }.
\end{equation}
This equation is analogous to Eq.~(\ref{eq:n_mode_exp_final}). Note, however, that in this case we use
subscripts instead of superscripts, while bar-functions are defined using underbars.

The total energy of a molecular system can, thereafter in the many-body expansion, be given as a sum over FC contributions,
\begin{equation}
    E(\mathbf{z}) =   \sum_{ \mathbf{f}_{j^{\prime}} \in \mathrm{FCR}  }  p_{ \mathbf{f}_{j^{\prime}}}   E_{ \mathbf{f}_{j^{\prime}} }.
\end{equation}
The coefficients $p_{ \mathbf{f}_{j^{\prime}}}$ thereby serve the purpose of weights for the corresponding energy terms $E_{ \mathbf{f}_{j^{\prime}}}$ 
in the overall expansion.
As was shown in Ref.~\cite{koenig2016}, these weights can be equal to zero for some energy contributions, 
especially if additional approximations are invoked.
For example, if a chain-like molecular system 
A--B--C--D is considered, 
the strongest interaction would probably be observed between neighboring fragments, e.g., A and B, B and C, etc. 
Therefore, it is often sufficient to include only those FCs, which are composed of neighboring fragments into the FCR. The corresponding
FCR that is closed on forming subsets and includes fragment interactions of up to second order, can be written as 
\begin{equation} \label{eq:FCR_complete}
    \mathrm{FCR}^{\mathrm{2F,NB}} = \{ \{ \}, \{ A \}, \{ B \}, \{ C \}, \{ D \}, \{ A,B  \}, \{ B,C  \}, \{ C,D  \}   \},  
\end{equation}
where FCs such as $\{A,C\}$, $\{A,D\}$, etc.\ are excluded from the FCR, as they are not composed of neighboring fragments.
In this expression, we use superscript NB to show that only neighboring interactions are considered. 
This results in 
energy contributions $E_{ \mathbf{f}_{j^{\prime}} }$ for FCs $\{A\}$ and $\{D\}$
cancel completely ($p_{\{\mathrm{A}\}}=0$ and $p_{\{\mathrm{D}\}}=0$; for more
detail, see Ref.~\cite{koenig2016}) leading to the effective FCR, 
\begin{equation} \label{eq:FCR_effective}
    \mathrm{FCR}^{\mathrm{2F,NB,eff}} = \{ \{ B \}, \{ C \}, \{ A,B  \}, \{ B,C  \}, \{ C,D  \}   \}. 
\end{equation}
The use of effective FCRs results in the same value of the total energy $E(\mathbf{z})$, but at a decreased 
computational cost compared to the use of a complete FCR.  
In the case of an arbitrary long chain-like molecular system F$_1$--F$_2$--$\cdots$--F$_N$, 
within the neighbor coupling approximation, 
coefficients $p_{ \mathbf{f}_j}^{l\mathrm{F, NB}}$ can be determined as the following~\cite{koenig2016}, 
\begin{equation} \label{eq:p_coeffs}
    p_{ \mathbf{f}_j}^{l\mathrm{F, NB}}=
    \begin{cases}
        1, & \forall j = l. \\
        -1, & \forall j = (l-1) \,\, \land  \,\, \mathbf{f}_j\, \mathrm{connected}  \,\,  \land   \,\, \mathbf{f}_j \cap \{ 1, N \} = \emptyset.  \\
        0, & \text{else}.
    \end{cases}
\end{equation}
Here, $j$ is the order of FC $\mathbf{f}_j$ and $l$ is the highest order of an FC in the FCR.
Accordingly, all FCs $\mathbf{f}_j$, for which the order deviates from the highest truncation level $l$ by more than one, 
cancel completely and can be omitted in the many-body expansion.
FCs with the order $j$ being equal to $(l-1)$ contribute to the FCR only if they are composed of directly 
connected fragments such as F$_i$--F$_{i+1}$--F$_{i+2}$ and do not include terminal fragments 
$F_1$ and $F_N$. Finally, contributions of all 
neighboring FCs with $j=l$ should always be included into the FCR.

The computational savings that can be expected from the use of these 
effective FCRs can easily be estimated. To this end, we consider 
a chain-like molecular system composed of $N_{\mathrm{frag}}$ chemically equivalent non-covalently bonded 
fragments of $N_{\mathrm{nuc}}$ nuclei.  
Assuming that the electronic structure method has a formal scaling of $O(N_{\mathrm{nuc}}^s)$, we can estimate
the cost of SP calculations for FCs composed of one and two fragments, as  
being proportional to $N_{\mathrm{nuc}}^s$ and $(2 N_{\mathrm{nuc}})^s$, respectively.
The computational cost of the approximate energy $E^{(l)} (\mathbf{z})$ calculated using the complete FCR 
can be estimated by 
\begin{equation}\label{eq:cost_of_E_complete}
    \sum_{i=1}^{l}\binom{N_{\mathrm{frag}}}{i} (i N_{\mathrm{nuc}})^s.
\end{equation}
Here, the sum runs over all FC orders up to the highest truncation order $l$, whereas
the binomial coefficient $\binom{N_{\mathrm{frag}}}{i}$ is equal to the number of 
FCs of order $i$ in the complete FCR.
Using Eq.~(\ref{eq:p_coeffs}) we can construct an effective FCR and omit low-order FCs.
This leads to the computational cost of $E^{(l)} (\mathbf{z})$ being proportional to  
\begin{equation} \label{eq:cost_of_E_effective}
    (N_{\mathrm{frag}} -l ) [(l-1) N_{\mathrm{nuc}}]^s + (N_{\mathrm{frag}}-l+1) (l N_{\mathrm{nuc}})^s.
\end{equation}
Therefore, the computational savings of using an effective FCR 
compared to a complete FCR
might be considerable 
for long chain-like molecular structures,
especially if high order truncation levels in the many-body expansion are needed.    
For example, for $N_{\mathrm{frag}}= 10$, $N_{\mathrm{nuc}}=5$, and $s=3$, the
use of an effective FCR results in SP calculations of $E^{(2)} (\mathbf{z})$,
$E^{(3)} (\mathbf{z})$, and $E^{(4)} (\mathbf{z})$
that are about 4.6, 13.3, and 28.0 times faster, respectively, compared to SP calculations employing a complete FCR. 
The leading terms in Eqs.~(\ref{eq:cost_of_E_complete}) and (\ref{eq:cost_of_E_effective}), 
however, stay unchanged as they depend on the highest-order FCs.

\subsection{Double Incremental Expansion\label{sec:die}}

In order to combine Eqs.~(\ref{eq:n_mode_exp_final}) and (\ref{eq:many_body_exp_final})
and obtain a double incremental expansion of the PES, 
we first need to consider a cut-potential $V^{ \mathbf{m}_k}$ for a mode combination $\mathbf{m}_k$. Its relation to the total 
energy of a molecular system can trivially be established by the expression,
\begin{equation} \label{eq:connection_V_E}
    V^{ \mathbf{m}_k} = E^{\mathbf{m}_k} - E (\mathbf{r}_0)  = \Delta E^{\mathbf{m}_k}, 
\end{equation}
where $E (\mathbf{r}_0)$ is the total energy of the reference molecular structure and 
$E^{\mathbf{m}_k}$ is the total energy obtained for the coordinates $z(\{ \{q\}^{\mathbf{m}_k},  \{0 \}^{m\not\in\mathbf{m}_k} \}  )$.
The set of coordinates $ \{\{q\}^{\mathbf{m}_k},  \{0 \}^{m\not\in\mathbf{m}_k}\}$ contains only those non-zero contributions $q_i$, which are included into
the MC $\mathbf{m}_k$, while all remaining coordinates are equal to zero.
A coordinate transformation between $\mathbf{z}$ and $\mathbf{q}$ is given in Ref.~\cite{koenig2016} as 
\begin{equation}
    \mathbf{z} = \mathbf{r}_0 + \mathbf{M}^{\frac{1}{2}} \mathbf{Lq}. 
	\label{eq:x_q_transformation}
\end{equation}
$\mathbf{r}_0$ is a matrix containing rectilinear coordinates of a reference configuration,  
$\mathbf{L}$ is orthogonal transformation matrix, and $\mathbf{M}$ is a 
diagonal $3 N_{\mathrm{nuc}}\times 3 N_{\mathrm{nuc}}$ matrix composed of diagonal sub-matrices of the form $m_i \mathds{1}$, 
where $m_i$ is a nuclear mass and $\mathds{1}$  is the $3\times 3$ identity matrix.
Incremental expansion of terms $E^{\mathbf{m}_k}$ and $E (\mathbf{r}_0)$ in Eq.~(\ref{eq:connection_V_E}) results in
\begin{equation} \label{eq:connection_V_E_expanded}
    V^{ \mathbf{m}_k} = \sum_{ \mathbf{f}_j \in \mathrm{FCR}  } [  \underline{E}^{\mathbf{m}_k}_{\mathbf{f}_j} - \underline{E}_{\mathbf{f}_j} (\mathbf{r}_{0, \mathbf{f}_j})                 ]
    = \sum_{ \mathbf{f}_j \in \mathrm{FCR}  } \Delta \underline{E}^{\mathbf{m}_k}_{\mathbf{f}_j}.
\end{equation}
Here, $\mathbf{r}_{0, \mathbf{f}_j}$ is the reference structure of the FC $\mathbf{f}_j$. 
The resulting bar-function $\Delta \underline{E}^{\mathbf{m}_k}_{\mathbf{f}_j}$ can then be re-written in terms of  
cut-functions similar to Eq.~(\ref{eq:many_body_exp_final}) such that  
\begin{equation}
    \Delta \underline{E}^{\mathbf{m}_k}_{\mathbf{f}_j} = \sum_{ \substack{
        \mathrm{\bf f}_{j^{\prime}} \subseteq  \mathrm{\bf f}_j   \\
        \mathrm{\bf f}_{j^{\prime}} \in \mathrm{FCR}  }}  (-1)^{j-j^{\prime}}   \Delta E^{\mathbf{m}_k}_{\mathbf{f}_{j^{\prime}}}, 
\end{equation}
where the FCR is assumed to be closed on forming subsets.
Finally, combining this expression with Eqs.~(\ref{eq:n_mode_exp_final}) and (\ref{eq:connection_V_E_expanded}), 
we obtain a double incremental expansion for the PES, 
\begin{equation}
    V^{(n,l)}(\mathbf{q}) =  \sum_{ \mathbf{m}_k \in \mathrm{MCR}  } \sum_{ \substack{
        \mathrm{\bf m}_{k^{\prime}} \subseteq  \mathrm{\bf m}_k   \\
        \mathrm{\bf m}_{k^{\prime}} \in \mathrm{MCR}  }}  (-1)^{k-k^{\prime}}
        \sum_{ \mathbf{f}_j \in \mathrm{FCR}  }
        \sum_{ \substack{
        \mathrm{\bf f}_{j^{\prime}} \subseteq  \mathrm{\bf f}_j   \\
        \mathrm{\bf f}_{j^{\prime}} \in \mathrm{FCR}  }}  (-1)^{j-j^{\prime}} 
        \Delta E^{\mathbf{m}_{k^{\prime}}}_{\mathbf{f}_{j^{\prime}}}.
\end{equation}
The approximate $n$M$l$F PES $V^{(n,l)}(\mathbf{q})$ is only equivalent to the exact PES $V(\mathbf{q})$ if truncation orders $n$ and $l$ are equal to  
the number of modes $M$ and the number of molecular fragments $N_{\mathrm{frag}}$, respectively.

Large computational savings can be obtained if the PES construction based on a double incremental expansion is combined with the 
use of strictly localized vibrational coordinates. 
This is due to the fact that only mode combinations, in which all modes move the 
 atoms in a certain fragment or FC have a non-zero contribution to the 
 PES energy surface of this 
 fragment or FC.  
This means all other terms can be 
 omitted in the double incremental expansion, which leads to 
 major computational savings 
(for more discussion on this topic, see Ref.~\cite{koenig2016}).
In this work, we apply purely vibrational semi-local FALCON coordinates~\cite{koenig2016_falcon} 
obtained by subspace diagonalization of the mass-weighted Hessian matrix of the entire molecular system.
The resulting coordinates have a well-defined spatial character and can be divided into two types:
i) coordinates that are strictly localized at a particular fragment and ii) coordinates that span multiple 
fragments. Following the notation used in Refs.~\cite{koenig2016_falcon,koenig2016,madsen2018}, we will denote these as 
intra-fragment (INTRA) and inter-connecting (IC) coordinates, respectively. Note that the approach in which the 
double incremental expansion is combined with the use of FALCON coordinates is generally referred to as DIF. 

\subsection{Adaptive Density-Guided Approach\label{sec:adga}}

The DIF methodology, as introduced in the previous section, is in this work combined with the ADGA~\cite{sparta2009} 
for construction of SP grids with which to represent the PES. The ADGA is an iterative 
 procedure that can determine grid granularity and dimensions without any specialized prior knowledge of the 
molecular system under investigation. This represents a clear advantage over static grid approaches, 
where the grid is predefined with individual SPs equidistantly placed. Note that all SPs, 
as determined to be important in the local sampling of the PES in the ADGA, are calculated via external 
electronic structure programs. A linear fit to the resulting grid is carried out, in order to obtain 
an analytic representation of the individual bar potentials, which can then be combined into the PES.

The grid points are generated via displacements from the reference point structure
in the ADGA, based on the relation given in Eq.~(\ref{eq:x_q_transformation}), i.e.,\ 
Cartesian coordinates at a displaced structure are given as  
\begin{equation} \label{eq:cart_disp_coord}
	\mathbf{z}^{\mathbf{m}_k}_{i^{\mathbf{m}_k}} =  \mathbf{r}_{0} + \sum_{m \in  \mathbf{m}_k} \mathbf{d}^m_{i^{ \mathbf{m}_k}} .
\end{equation}
This means that displacements along each vibrational mode $m \in \mathbf{m}_k$ is carried out, in order to obtain 
the particular nuclear arrangement $\mathrm{z}^{\mathbf{m}_k}_{i^{\mathbf{m}_k}}$ 
for each SP $i^{\mathbf{m}_k}$. Displacements along rectilinear vibrational 
    coordinates, as the FALCON coordinates, can be defined as
\begin{equation}\label{eq:cart_disp_coord_2}
\mathbf{d}^m_{i^{\mathbf{m}_k}} = \mathbf{M}^{-\frac{1}{2}} \mathbf{l}_m \eta^m_{i^{\mathbf{m}_k}} \Lambda_m .
\end{equation} 
The column vector $\mathbf{l}_m$ should be understood as the $m$th column of the transformation matrix $\mathbf{L}$. 
The fractional displacements $\eta^m_{i^{\mathbf{m}_k}} \in [-1,1]$ belong to the set of rational
numbers, while the scale factor $\Lambda_m$ is given in terms of the harmonic oscillator (HO) quantum 
number $v$ and harmonic angular frequency $\omega_m$ as 
\begin{equation}
\Lambda_m = N_{\textup{IBP}} \sqrt{\frac{2 \hbar}{\omega_m} \left( v + \frac{1}{2} \right)} .
\end{equation}
The initial boundary points for each mode are placed at fractional positions $\eta^m_1 = - \frac{1}{N_{\textup{IBP}}}$ 
and $\eta^m_2 = \frac{1}{N_{\textup{IBP}}}$. This means that the first two SPs for each mode are always situated at positions 
that correspond to the classical turning points for an HO. 
The interval defined between these two points is sometimes referred to, as the initially spanned space.

The ADGA is guided by means of vibrational self-consistent field (VSCF) 
calculations, in order to obtain the one-mode average vibrational density, 
\begin{equation}
    \rho^{\mathrm{ave}}_{\mathrm{iter}} (q_{m_i}) = \frac{1}{N^{m_i}_{\mathrm{modal}}} \sum^{N^{m_i}_{\mathrm{modal}}}_{s^{m_i} =1} 
    | \varphi_{s^{m_i}}^{m_i}  ( q_{m_i}  ) |^2 ,
\end{equation}
where $\varphi_{s^{m_i}}^{m_i}  ( q_{m_i}  )$ are orthonormal one-mode wave functions, 
usually denoted modals, while $N^{m_i}_{\mathrm{modal}}$ is the number of modals to be 
determined for each vibrational mode $q_{m_i}$ and $s_i$ defines which modal is occupied. 
The average vibrational density is part of the energy-like quantity, which in the one-mode case reads, 
\begin{equation} 
    \Omega_{\mathrm{iter}}^{m_i} = \int_{I_{m_i}} \rho^{\mathrm{ave}}_{\mathrm{iter}} (q_{m_i}) 
    V^{m_i}_{\mathrm{iter}} (q_{m_i}) \mathrm{d} q_{m_i}.
    \label{eq:adga_int}
\end{equation}
$ \Omega_{\mathrm{iter}}^{m_i}$ is evaluated by the ADGA in each iteration ($\mathrm{iter}$) and
 for each interval $I_{m_i}$, defined by consecutive pairs of SPs in the grid.
$V^{m_i}_{\mathrm{iter}} (q_{m_i})$ is the one-mode potential in the iteration $\mathrm{iter}$.
 This is done to investigate if substantial changes are found between the previous and 
current iterations~\cite{klinting2018}. If a substantial relative change is found 
in $\Omega_{\mathrm{iter}}^{m_i}$ between iterations in the ADGA, then the corresponding interval 
is subdivided by the inclusion of a new SP, unless the absolute change is small. 
Furthermore, if a significant amount of the average vibrational density is found outside of 
the current grid, then the grid is extended. These three aspects of the ADGA are 
controlled via convergence criteria with associated thresholds $\epsilon_{\textup{rel}}$, 
$\epsilon_{\textup{abs}}$ and $\epsilon_{\rho}$, which are further detailed in 
Refs.~\cite{sparta2009,klinting2018}. This process is continued until convergence 
is established for all intervals defined for one-mode MC grids, whereafter the process continues 
analogously for 
two-mode MC grids and so forth, until convergence at the desired mode combination level is 
reached. Note that the average vibrational density is only calculated for the non-coupled (1M) 
PES in the ADGA and a direct product of these are used for higher-dimensional 
cases, i.e., to generate $ \rho^{\mathrm{ave},\mathbf{m}_k}_{\mathrm{iter}}$, which 
 are used in the ADGA procedures for the higher-dimensional PES cuts  $\bar{V}^{\mathbf{m}_k}_{\mathrm{iter}}$~\cite{toffoli2011}, 
 where we have omitted the coordinate dependence for clarity. 
 Moreover, grid boundaries remain fixed for higher-order MC grids, 
after their determination for the corresponding one-mode MC grids.

\subsection{Combining the ADGA and Double Incremental Expansion}

An algorithm for the combined DIF-ADGA methodology is presented in Alg.~\ref{algo:dif_adga}.
Its distinct difference from the standard ADGA lies in the 
necessity to construct and later combine fragment PESs. 
Similar to the standard ADGA, we start the iterative procedure from the lowest order 
PES ($k=1$) and proceed to higher orders upon convergence.
For any particular FC $\mathbf{f}_{j^{\prime}}$,
we consider only those MCs $\mathbf{m}_{k}$, which are composed exclusively of modes moving the respective fragments.  
For each such combination of  FC $\mathbf{f}_{j^{\prime}}$ and 
MCs $\mathbf{m}_{k}$,
a set of displaced molecular geometries and the corresponding energies $\Delta E_{\mathbf{f}_{j^{\prime}}}^{\mathbf{m}_{k}} $
are determined.
These energies are summed up over all MCs $\mathrm{\bf m}_{k^{\prime}}$ as
\begin{equation} \label{eq:bar_pot_algo_33}
    \bar{V}^{ \mathrm{\bf m}_{k}   }_{ \mathrm{\bf f}_{j^{\prime}}} = 
    \sum_{ \substack{
        \mathrm{\bf m}_{k^{\prime}} \subseteq  \mathrm{\bf m}_k   \\
        \mathrm{\bf m}_{k^{\prime}} \in \mathrm{MCR}  }}  (-1)^{k-k^{\prime}}
    \Delta E_{\mathbf{f}_{j^{\prime}}}^{\mathbf{m}_{k^{\prime}}} 
\end{equation}
and fitted to obtain an analytical representation of the bar-potential 
$\bar{V}^{ \mathrm{\bf m}_{k}   }_{ \mathrm{\bf f}_{j^{\prime}}} $. 
This is repeated for all FCs $\mathbf{f}_{j^{\prime}}$ and MCs $\mathbf{m}_{k^{\prime}}$.
The resulting fitted bar-potentials are then combined for all FCs such that
\begin{equation} \label{eq:bar_pot_algo_34}
    \bar{V}^{ \mathrm{\bf m}_{k}   }_{\mathrm{iter}} = 
        \sum_{ \mathbf{f}_j \in \mathrm{FCR}  }
        \sum_{ \substack{
        \mathrm{\bf f}_{j^{\prime}} \subseteq  \mathrm{\bf f}_j   \\
        \mathrm{\bf f}_{j^{\prime}} \in \mathrm{FCR}  }}  (-1)^{j-j^{\prime}} 
        \bar{V}^{ \mathrm{\bf m}_{k}   }_{ \mathrm{\bf f}_{j^{\prime}}}.
\end{equation}
If the condition $k=1$ is satisfied, these bar-functions are further combined into the 1M PES $V^{(1,l)}$ and used to 
obtain the average vibrational densities $\rho^{\mathrm{ave},{\mathbf{m}_{k}}}_{\mathrm{iter}}$ via VSCF calculations and determine the correct one-mode boundaries.
For $k>1$, the averaged vibrational density is constructed as a direct product of one-mode densities, 
while one-mode grid boundaries are used to determine also the 
 multi-dimensional grid boundaries. 
The convergence of the ADGA is determined with respect to the densities $\rho^{\mathrm{ave},{\mathbf{m}_{k}}}_{\mathrm{iter}}$
and potentials $\bar{V}^{\mathbf{m}_{k}}_{\mathrm{iter}}$ on intervals $I^{\mathbf{m}_{k}}$, all three referring to the full system as opposed to subsystems.
If some $\bar{V}^{\mathbf{m}_{k}}_{\mathrm{iter}}$ is not converged, new displaced structures and new SPs are calculated.
The procedure is repeated until convergence for all $n$ levels is achieved. The resulting 
$\bar{V}^{ \mathrm{\bf m}_{k}}_{\mathrm{iter}}$ potentials are then combined to form the total 
$n$M$l$F molecular PES $V^{(n,l)}$.

\begin{algorithm}[]
    \For{\textnormal{MC levels} $k = 1, 2, \dots, n$}
    {
    \For{\textnormal{ADGA iterations}}
    {
        \For{\textnormal{FCs} $\mathbf{f}_{j} \in \textnormal{FCR}$}
        { 
            \For{\textnormal{those MCs} $\mathbf{m}_{k} \in$ \textnormal{MCR, which modes all move atoms in} $\mathbf{f}_{j}$}
            {
                Generate SPs \\
                \Indp 1. Obtain Cartesian coordinates [Eqs.~(\ref{eq:cart_disp_coord}) and (\ref{eq:cart_disp_coord_2})] \\
                 2. Calculate energy $\Delta E_{\mathbf{f}_{j}}^{\mathbf{m}_{k}} $ \\
                 3. Generate $\bar{V}^{ \mathrm{\bf m}_{k}   }_{ \mathrm{\bf f}_{j^{\prime}}}  $ SPs [Eq.~(\ref{eq:bar_pot_algo_33})] \\ 
                \Indm 
                Fit to grid of SPs $\bar{V}^{ \mathrm{\bf m}_{k}   }_{ \mathrm{\bf f}_{j^{\prime}}}  $ \\
            }
        }

        \For{\textnormal{FCs} $\mathbf{f}_{j} \in \textnormal{FCR}$}
        { 
            \For{\textnormal{FCs} $\mathbf{f}_{j^{\prime}} \in \textnormal{FCR}$, $\mathbf{f}_{j^{\prime}} \subseteq  \mathbf{f}_{j}$}
            {
                Combine $\bar{V}^{ \mathrm{\bf m}_{k}   }_{ \mathrm{\bf f}_{j^{\prime}}}  $ into  $\bar{V}^{ \mathrm{\bf m}_{k}   }_{\mathrm{iter}} $ [Eq.~(\ref{eq:bar_pot_algo_34})] 
            }
        }

        \If{$k=1$}
        {
            Combine $\bar{V}^{ \mathrm{\bf m}_{k}   }_{\mathrm{iter}} $ into $V^{(1,l)}_{\mathrm{iter}}$ \\
	        Calculate $\rho^{\mathrm{ave},{\mathbf{m}_{k}}}_{\mathrm{iter}}$  by running VSCF on $V^{(1,l)}_{\mathrm{iter}}$
	    }       

            \For{\textnormal{MCs} $\mathbf{m}_{k} \in \textnormal{MCR}$}
            {
	            Evaluate integrals 
                $\int_{I^{\mathbf{m}_{k}}} 
                    \rho^{\mathrm{ave},{\mathbf{m}_{k}}}_{\mathrm{iter}} 
                    \bar{V}^{\mathbf{m}_{k}}_{\mathrm{iter}}  
                    \mathrm{d} \mathbf{q}^{\mathbf{m}_{k}}$ 
                and $\int_{I^{\mathbf{m}_{k}}} 
                    \rho^{\mathrm{ave},{\mathbf{m}_{k}}}_{\mathrm{iter}} 
                    \mathrm{d} \mathbf{q}^{\mathbf{m}_{k}}$ 
                on supersystem grid intervals  $I^{\mathbf{m}_{k}}$ \\
                Check convergence for all $I^{\mathbf{m}_{k}}$ intervals \\
                
                \If{\textnormal{Not converged}}
                {
	                Define new displacements for next iteration \\
	            }                
            }
    }
   } 
    Combine all converged $V^{ \mathrm{\bf m}_{k}}$ into the final full $n$M$l$F molecular PES $V^{(n,l)}$ 
\caption{Sketch of the DIF-ADGA for constructing an $n$M surface. }
\label{algo:dif_adga}
\end{algorithm}

\newpage
\clearpage

\section{Computational Details \label{sec:comp_details}}

To validate the DIF-ADGA methodology, fundamental frequencies of the dicyclopropyl ketone, 
tetraphenyl, and hexaphenyl molecules were calculated. 
The molecular structure of dicyclopropyl ketone was optimized 
with the {\sc Orca} program~\cite{orca} using the method of Hartree--Fock 
with three corrections (HF-3c)~\cite{sure2013}.
Structures of tetraphenyl and hexaphenyl were provided by the authors of Refs.~\cite{koenig2016,madsen2018}. 
Vibrational FALCON coordinates~\cite{koenig2016_falcon}
similar to those reported in Refs.~\cite{koenig2016,madsen2018}
were generated for all three molecules with the 
Molecular Interactions Dynamics And Simulation Chemistry Program Package
({\sc MidasCpp})~\cite{midascpp}. 
The procedure was carried out in such a way that INTRA coordinates
are localized at the carbonyl and the two cyclopropyl groups for the case of dicyclopropyl ketone
and localized at each phenyl ring for tetraphenyl and hexaphenyl. 
The HF-3c approach was employed in calculations of the reduced Hessian matrices used in setting up the FALCON coordinates.
The resulting coordinates were used for PES construction at the HF-3c level of theory with i) a static grid approach~\cite{toffoli2007},
ii) a static grid approach employing DIF~\cite{koenig2016}, iii) the ADGA to grid construction~\cite{sparta2009,klinting2018}, 
and iv) the DIF-ADGA to grid construction.
The static grids 
were constructed using 
20 SPs for each one-mode MC grid and
10$\times$10 SPs for each two-mode MC grid with equidistant grid points.
The static grid boundaries were set to match the turning points for an HO of quantum number $v=10$.
In ADGA and DIF-ADGA calculations, a convergence threshold of $\epsilon_{\textup{abs}} = 1.0\times 10^{-6}$~a.u.\ was applied for 
changes in the integrated density times potential values, as seen in Eq.~(\ref{eq:adga_int}).
Other convergence criteria (see Refs.~\cite{sparta2009,klinting2018}) were set to their default values, 
$\epsilon_{\textup{rel}} = 1.0\times 10^{-2}$ and 
$\epsilon_{\rho} = 1.0\times 10^{-3}$.
Standard non-fragmented ADGA calculations were 
employed with the utilities of gradient guided basis set determination
and dynamic extension of grid boundaries~\cite{klinting2018}.
In these calculations, the initial HO turning point was defined by the quantum number $v=2$. 
Several different quantum numbers $v$ were tested for the DIF-ADGA and discussed in Sec.~S1 in the Supporting Information (SI).
The value of $v=8$ was found optimal for DIF-ADGA calculations and was used throughout this work.
Fragmentation of molecular structures in the DIF and DIF-ADGA was carried out 
in full analogy to the coordinate localization procedure and reported in Refs.~\cite{koenig2016,madsen2018}.
Note that the dangling bonds were capped with hydrogen atoms in the fragmentation scheme, as 
 outlined further in Ref.~~\cite{koenig2016}. 
The neighbor coupling approximation was invoked when generating the FCR for all three molecules.
For a detailed description of FCRs used for these test systems, see Fig.~S2, and Tab.~S1 in the SI.
The $n$-mode expansion and the many-body expansion were both truncated at the second order 
thereby determining the maximum sizes for the MCR and FCR.
The resulting PESs will, therefore, be referred to as $n$M Static or ADGA and $n$M$l$F DIF-Static or DIF-ADGA.
Analytical representation of the PESs were obtained from the linear fit using up to 12th order polynomial fit-basis functions.
The fitted PESs were used in VSCF calculations~\cite{bowman1978,gerber1979,christiansen2004,hansen2010} 
to determine
fundamental excitation energies, for which 80 B-splines (per mode) of order 10 were employed as the primitive basis for 
representing the vibrational wave function~\cite{toffoli2011}. 
This number of B-splines was found 
sufficient to obtain converged results for fundamental excitation energies.

The computational time required for PES evaluations was estimated using HF-3c, the 
Kohn--Sham Density Functional Theory (KS-DFT), and the second-order M{\o}ller--Plesset perturbation theory (MP2).
In the case of KS-DFT, the BP86 
exchange--correlation functional~\cite{beck1988,perd1986} was employed.
KS-DFT and MP2 approaches were used in conjunction with the resolution-of-the-identity approximation,
the Ahlrichs split valence basis set of double-$\zeta$ quality with one set of 
polarization functions (SVP)~\cite{schaefer1992,schaefer1994}, and 
the auxiliary Coulomb fitting SV/J basis set~\cite{eichkorn1995,eichkorn1997}.  
All PES computations were carried out in parallel with each SP executed on a single core using nodes 
Intel Xeon X5650 @ 2.7 GHz/49Gb with 12 cores in total.  
The KS-DFT and MP2 approaches were used to calculate small regions of PESs for   
the tetraphenyl and hexaphenyl molecules as well as for their FCs.
The SP computational cost for each molecule and FC was computed by taking an average 
of corresponding 100 SP run times. 
For HF-3c, SP times (averaged over entire PESs) were provided by authors of Ref.~\cite{madsen2018}.
The times required for the total HF-3c, KS-DFT, and MP2 PES evaluations were then
computed by multiplying these average SP costs by numbers of SPs obtained in HF-3c PES calculations and,
in case of DIF computations, summing up over all FCs.

Calculations of the fundamental excitation energies of 
the (3E,5E,7E,9E,11E,13E,15E,17E)-icosa-1,3,5,7,9,11,13,15,17,19-decaene molecule, 
in the following referred to as icosadecaene, were carried out in order to further investigate the convergence of the many-body expansion. 
The fundamental frequencies were determined via VSCF calculations based on PESs constructed by 
the ADGA or DIF-ADGA using the same computational setup as described above.
Two different types of fragmentations and
FALCON procedures were used. First, we considered 
$\ce{-CH=CH-}$ units as separate fragments and local FALCON coordinates at these, thus, dividing the 
entire molecular system into 10 subsystems. Secondly, we
combined pairs of such neighboring units and generated a different set of FALCON 
coordinates for the resulting 5 subsystems. 
1M$l$F PESs with $f=1,2,\dots,9$ were constucted via the DIF-ADGA for the former case, while 
1M$l$F PESs with $f=1,2,\dots,4$ were constructed for the latter case.
The effective FCRs are explicitly given in Fig.~S2 and Tab.~S2 of the SI.

\section{Results \label{sec:results}}

In this section, we first validate the DIF-ADGA and compare it with other methods of grid construction 
in Sec.~\ref{sec:validation}.
Secondly, in Sec.~\ref{sec:convergence} the convergence of the double incremental expansion is 
investigated with respect to the truncation order $l$ in the associated many-body expansion.

\subsection{Validation of the DIF-ADGA \label{sec:validation}}

\begin{figure}[h!]  
\centering 
\includegraphics[width=0.90\linewidth]{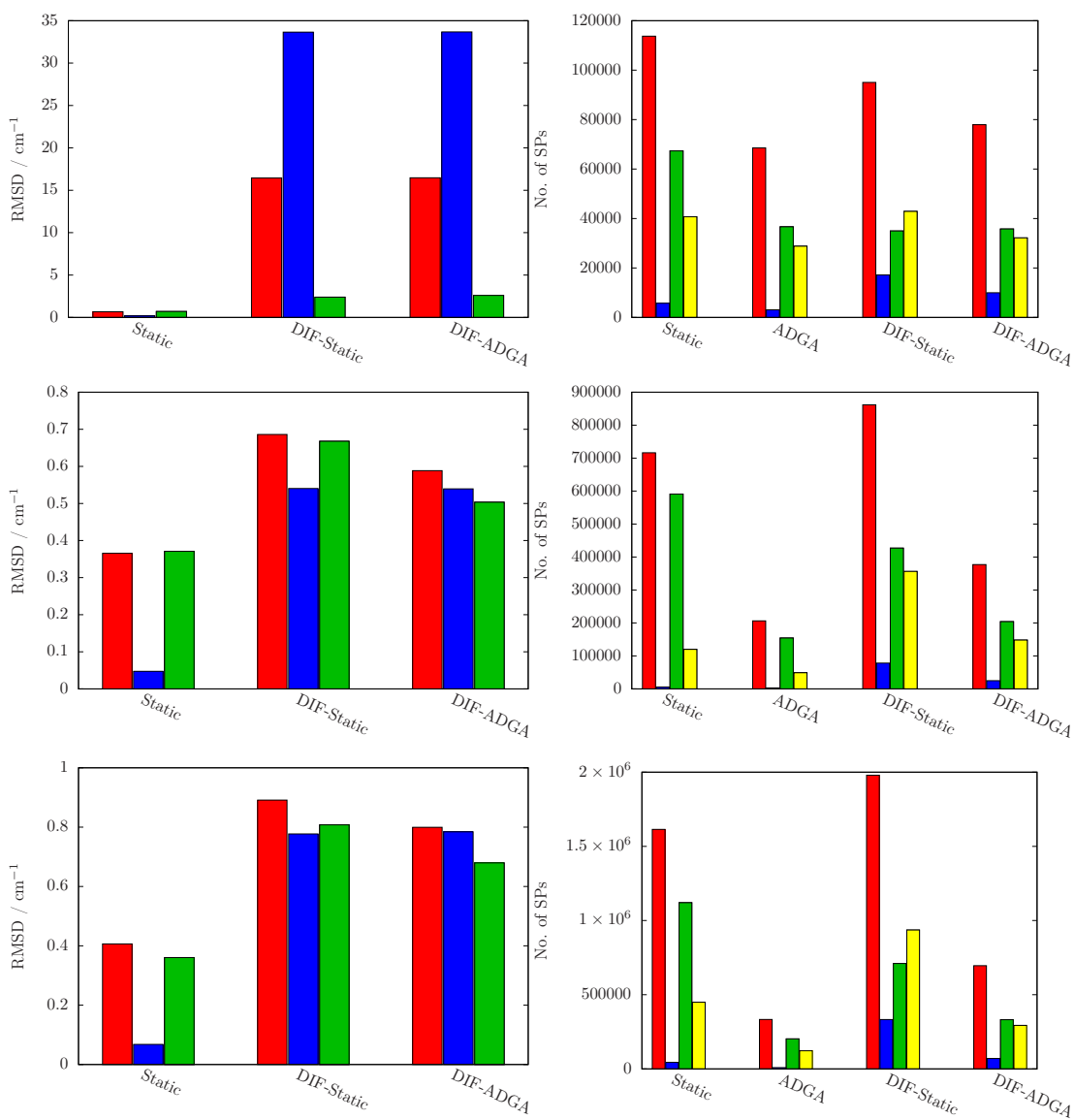}  
    \caption{Results of 2M PESs generations for dicyclopropyl ketone, 
    tetraphenyl, and hexaphenyl (from top to bottom, respectively).
    RMSDs of VSCF fundamental 
    frequencies from supersystem ADGA are shown on the left, while numbers of SPs used for PES 
    generations are on the right. 
    Red bars show RMSDs and SPs for the entire set of vibrational modes, whereas blue and green bars 
    represent values calculated for IC and INTRA modes, respectively. SPs calculated for 
    IC--INTRA coupling potentials are shown in yellow.} \label{fig:2M_case}
\end{figure}

To validate our newly proposed methodology, we performed
computations using four different schemes for PES construction, namely  
the non-fragmented static grid construction approach and the ADGA as well as the fragmentation-based DIF-Static and DIF-ADGA algorithm.
The results of these calculations for 2M and 2M2F PESs are presented in Fig.~\ref{fig:2M_case}. 
As one can see from Fig.~\ref{fig:2M_case} (left), the static grid approach and ADGA perform very similar with
root-mean square deviations (RMSDs) of about 
0.4 cm$^{-1}$ for all three molecules in non-fragmented cases. 
Interestingly, the largest deviations are found for the total set of vibrational modes and for INTRA vibrations.
Fragmentation errors in fundamental excitations from the DIF-Static are system dependent and are equal to about 16.4 cm$^{-1}$, 
0.7 cm$^{-1}$, and 0.9 cm$^{-1}$ for the dicyclopropyl ketone, tetraphenyl, and hexaphenyl molecule, respectively.
These results are in very good agreement with those previously reported in Refs.~\cite{koenig2016,madsen2018}. 
In these works, the RMSDs in fundamental excitations caused by fragmentation were found to be 20.15, 0.56, 0.75 cm$^{-1}$, respectively.
We, hence, reproduced the RMSD values for the tetraphenyl and hexaphenyl molecule
from the previous studies. 
The small deviations for tetra- and hexaphenyl are due to slightly different setups of the 
 static grid, while
 the larger difference between the obtained RMSDs 
calculated for dicyclopropyl ketone can be explained by the use of different electronic structure 
 methods: The authors of 
Ref.~\cite{madsen2018} applied MP2 in both FALCON coordinate optimization and
SP calculations, while we used the HF-3c approach for these purposes.
The resulting coordinates and PES landscapes calculated in this work and in Ref.~\cite{madsen2018}, therefore, differ to some extent.
DIF-ADGA fragmentation errors were found to be very similar to those from DIF-Static with differences in RMSDs of about 
0.1 cm$^{-1}$ in all cases.
Therefore, the DIF-Static scheme and the DIF-ADGA lead to essentially the same results.
Very large DIF-ADGA and DIF-Static RMSDs compared to the full molecule are found for dicyclopropyl ketone and can probably be explained by comparably small sizes of molecular fragments and
strong interactions between them. In these cases, almost eight times larger deviations are observed for IC modes
than for INTRA vibrations. The latter are, therefore, less affected by the fragmentation.
Larger molecular fragments in tetraphenyl and hexaphenyl and a rather weak interaction between them results
in much smaller DIF-ADGA and DIF-Static RMSDs for the total set of vibrational modes and a small difference in RMSDs of IC and INTRA vibrations.
Thus, the accuracy of DIF-based schemes is, as expected, strongly dependent on sizes and types of chosen molecular fragments.
One might expect that fragmentation errors will always be very large, if the interactions are strong. 
However, as will be shown in Sec.~\ref{sec:convergence} results of very high accuracy can be achieved even in 
very unfavorable cases for fragmentation by including higher-order FCs in the FCR.

As can be seen from Fig.~\ref{fig:2M_case} (right), 
PES construction using the ADGA always requires less SPs than the static grid approach. 
Almost twice the number of SPs is needed for the dicyclopropyl ketone PES, when using the static grid approach
instead of the ADGA, with the current settings. 
An even larger reduction of SPs is observed for tetraphenyl and hexaphenyl.
For these two molecules, the numbers of SPs used in the ADGA are 3.5 and 4.8 times smaller than for the static grid approach. 
In both the static grid approach and ADGA calculations, the total number of SPs is largely dominated by
contributions from INTRA modes, while the smallest number of SPs is required for IC modes.
This can be explained by the fact that the number of IC modes is smaller or equal to $(N-1)6$, where $N$ is the 
number of fragments, and, therefore, is always much smaller than the total number of vibrational modes~\cite{koenig2016}.
The use of the fragmentation-based DIF-Static approach leads to an increase in the number of SPs compared to the standard 
non-fragmented
ADGA scheme for all three molecules. 
For the cases of tetraphenyl and hexaphenyl, the number of SPs used in the DIF-Static is also larger than those obtained with the static grid.
This is due to the fact that multiple FCs need to be calculated, while using the DIF-Static approach (see Tab.~S1 in the SI).   
Also, because each IC mode spans more than one fragment, displacements along them should be computed for 
each FC that includes these fragments. 
The latter can also be seen from the greatly increased numbers of SPs for IC modes and IC--INTRA couplings 
in Fig.~\ref{fig:2M_case} (right).
It should, however, be noted that due to fragmentation the resulting SPs are less 
expensive than those calculated with the non-fragmented static grid approach and ADGA.

As can be seen from Fig.~\ref{fig:2M_case} (right), the DIF-ADGA reduces the number of SPs required for PES construction for all three molecules 
compared to the DIF-Static scheme.  
This reduction is rather modest for the dicyclopropyl ketone molecule and reaches only about 
18\% of the total number of SPs. For tetraphenyl and hexaphenyl, computational savings are much larger and the number of SPs calculated 
with DIF-ADGA are 2.3 and 2.8 times smaller than those required for DIF-Static PESs.

\begin{figure}[h!]  
\centering 
\includegraphics[width=0.60\linewidth]{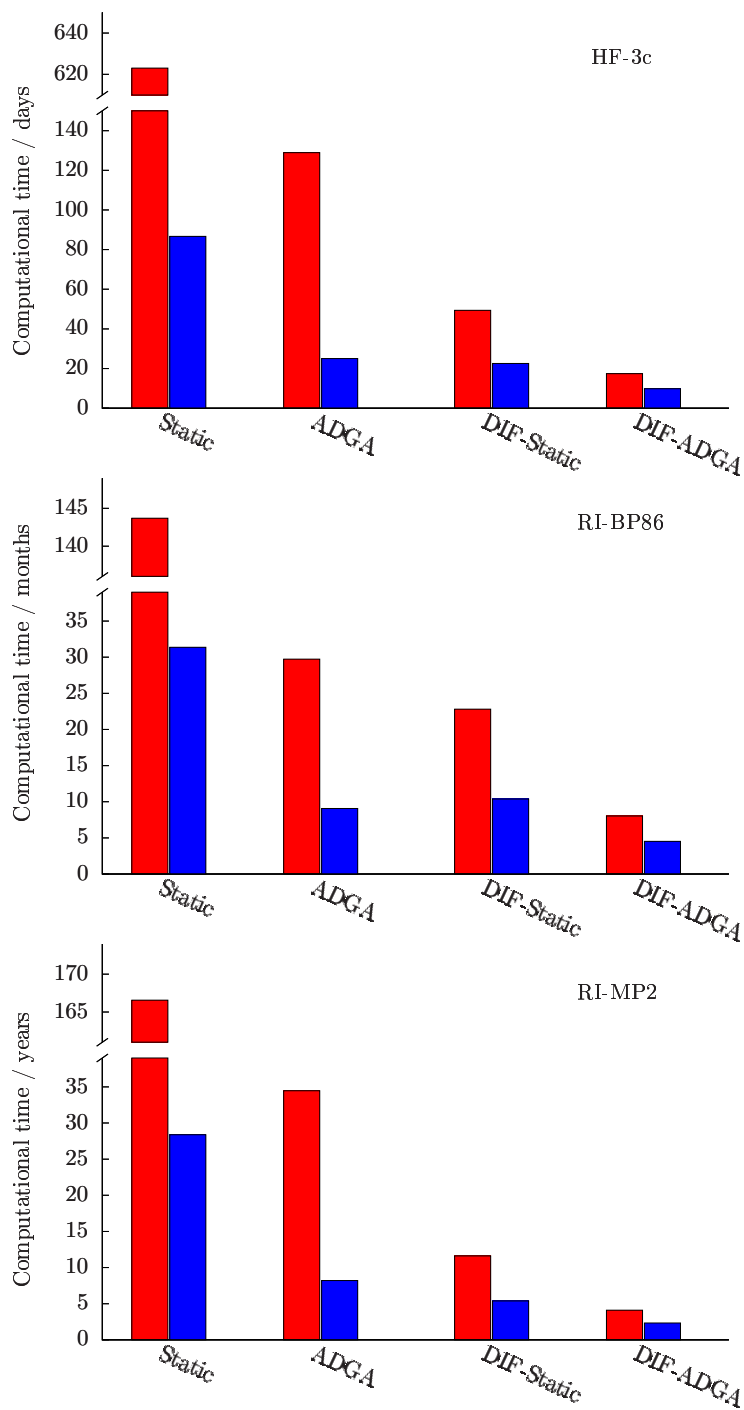}  
    \caption{Estimates for the accumulated time of all SPs required for PES constructions using the
    HF-3c (top), RI-BP86 (middle), and RI-MP2 (bottom) electronic structure methods.
    Results are shown for
    the hexaphenyl (red) and tetraphenyl (blue) molecules.} \label{fig:dft_timing}
\end{figure}

To further assess computational savings of the DIF-ADGA compared to DIF-Static, we estimated the time required 
for HF-3c, RI-BP86, and RI-MP2 PES constructions, as described in Sec.~\ref{sec:comp_details}.
For this, we assumed the same number of SP calculations for the use of RI-BP86 and RI-MP2
 that was required for the HF-3c PES.
 Note, however, that the use of different electronic structure methods with the ADGA leads to 
a different number of SP required for a PES generation. 
The provided timings are, therefore, rather crude estimates.   
We further note that the presented timings refer to the total accumulated 
costs  
 of all single-point calculations required in the 
 PES constructions conducted on a single core, where overheads such as the VSCF calculations required in the ADGA and fittings are not included. These, however, 
 are typically only very small fractions of the total cost.
In practice, the computations are of course done in parallel. 
The final estimates are shown in Fig.~\ref{fig:dft_timing}.
As one can see from Fig.~\ref{fig:dft_timing} (top), 
it requires about 620 and 
90 days to calculate 2M PES of the hexaphenyl
and tetraphenyl molecules, respectively, while applying the static grid approach and the HF-3c 
electronic structure method (on a single core). 
The use of the ADGA drastically decreases the 
computation cost by 3--5 times for both molecules. 
Even larger computational savings can be reached for hexaphenyl with the DIF-Static approach, whereas for the smaller tetraphenyl molecule
computational costs of the supersystem ADGA and DIF-Static are comparable in magnitude. 
The difference between hexaphenyl and tetraphenyl demonstrates that as the systems become larger, the advantage of the DIF approach increases. 
This is true for both the static grid and ADGA cases. 
The DIF-ADGA allows for 2--3 times less expensive PES computations compared to the DIF-Static approach.
Recalling that DIF-static and DIF-ADGA provided similar accuracy this reduction is clearly 
very attractive. 

Very similar trends are obtained for RI-BP86 and RI-MP2, as can be seen from the middle and 
 bottom graphs in Fig.~\ref{fig:dft_timing}, respectively. The overall computational times are much higher, 
but similar reductions as before are obtained with both DIF and the ADGA.
The DIF-ADGA is still 
the fastest approach by a significant factor. 
As expected the overall gain obtained from replacing the supersystem 
static grid approach with the DIF-ADGA increases as we move towards 
larger molecules. 
Thus, the DIF-ADGA computation for tetraphenyl is 12 times faster than the supersystem static grid PES construction, 
 whereas the respective 
factor for the hexaphenyl molecule is equal to 41.

The above-presented computational savings demonstrate an advantage of the DIF-ADGA over the DIF-Static approach.
Thus, for the hexaphenyl and tetraphenyl molecules the use of DIF-ADGA led to PES calculations, 
which are faster by a factor of 2--3 
than those with DIF-Static. 
These factors, however, depend on 
the size and granularity of static grids and ADGA settings applied. 
Finer static grids and looser ADGA convergence criteria would show higher computational savings than those
demonstrated in this work. 
Although we consider the applied static grids and ADGA convergence criteria reasonable, we cannot claim that both methods converge to 
the very same point as systematic studies on convergence of grids for this set of molecules were not conducted.
However, considering how the computational cost of different approaches 
varies with the size of the molecule is likely more independent of the grid quality.
We therefore compare numbers of SPs and computational times for tetraphenyl against results for hexaphenyl obtained with the RI-MP2 electronic structure method and same grid quality.
In the ideal case of linear scaling in the total computational cost,
calculations of hexaphenyl are expected to take about 1.5 longer than those for tetraphenyl,
based on counting vibrational degrees of freedom in the two computations.
The supersystem RI-MP2 static grid 
calculations are far from linear scaling and show a factor of 5.9 
for the accumulated time of all hexaphenyl electronic structure computations relative to the tetraphenyl computations, 
while the number of SPs for hexaphenyl is about 2.3 times larger for tetraphenyl. 
These factors decrease considerably for ADGA calculations and approach 4.2 and 1.6 for the computational cost and the number of SP, 
respectively. Therefore, nearly-linear scaling behavior can be seen for the number of SPs, as has similarly been reported in a systematic scaling 
study for conjugated hydrocarbons in Ref.~\cite{hansen2010}.
DIF-Static and the DIF-ADGA show a further decrease in the total accumulated electronic structure computational cost 
(giving hexaphenyl/tetraphenyl factors of 2.2 for DIF-Static and 1.8 for DIF-ADGA), even if 
slightly larger fractions for the numbers of SPs are obtained due to the fragmentation (2.3 and 1.8, respectively). Thus, even if computational timings are subject to noise and
overhead, the observed close to linear scaling in both the overall computational cost of the DIF-ADGA and the numbers of SPs
is encouraging with respect to future work on constructing PESs for large molecular systems.

We, therefore, conclude that the 
DIF-ADGA is a valuable and robust methodology for PES constructions.  
It demonstrates considerable computational savings compared to other approaches applied and 
allows for a completely automized grid construction, while maintaining a 
high accuracy of the resulting PES. The computational savings are also expected to increase 
with the level of electronic structure theory.
The only modest increase in the number of SPs when using DIF-ADGA instead of supersystem approaches is, however, only possible when the 
neighbor coupling approximation is invoked. 
Calculations of high-order non-neighbor FCs within the DIF-ADGA 
would require the lower-order FCs to be included into the FCR as well and, therefore, would lead to
a larger increase in the number of SPs (see Sec.~\ref{sec:many_body}). 
In fact, the standard many-body expansion scales polynomially with respect to the order of expansion, whereas 
the use of the neighbor coupling approximation can lead to the linear scaling behavior
(see Ref.~\cite{koenig2016}).

\subsection{Convergence of Many-Body Expansions \label{sec:convergence}}

As was discussed in Sec.~\ref{sec:many_body}, the many-body expansion of Eq.~(\ref{eq:many_body_exp_long}) 
provides the exact value for the energy $E (\mathbf{z})$, if the truncation order 
$l$ is equal to the number of variables $N_{\mathrm{frag}}$, while the case of $l<N_{\mathrm{frag}}$ leads to an approximate 
treatment of $E (\mathbf{z})$. Calculating higher-order couplings, one can   
reach a desirable level of accuracy and establish an optimal trade-off between efficiency and 
computational cost. It is, however, not clear \textit{a priori} how many terms need to be considered 
and what accuracy can be expected for a chosen truncation level.
This problem is especially severe in the case of the double incremental expansions of PESs, where 
the exact PES is unknown and the computational cost of the next $l+1$ term might be 
prohibitively high. 
Moreover, in practical calculations the quality of approximate PESs 
can vary considerably depending on many different aspects such as 
the type and size of fragments, strength of interactions between them, etc. 
This has already been demonstrated in Sec.~\ref{sec:validation} 
using the dicyclopropyl ketone, tetraphenyl, and hexaphenyl molecules as examples.   
Thus, negligibly small fragmentation errors, i.e., below 1 cm$^{-1}$, were achieved 
for the latter two molecules already at the very low truncation order $l=2$, 
while the very same truncation order led to rather large RMSDs in the case of the 
first molecule. Therefore, a deeper insight into the convergence of double incremental expansions for
molecular systems featuring different chemical properties and fragmentation patterns 
is necessary and can give some indication on how to effectively construct FCRs and run these calculations.

\begin{figure}[h!]  
\centering 
\includegraphics[width=0.80\linewidth]{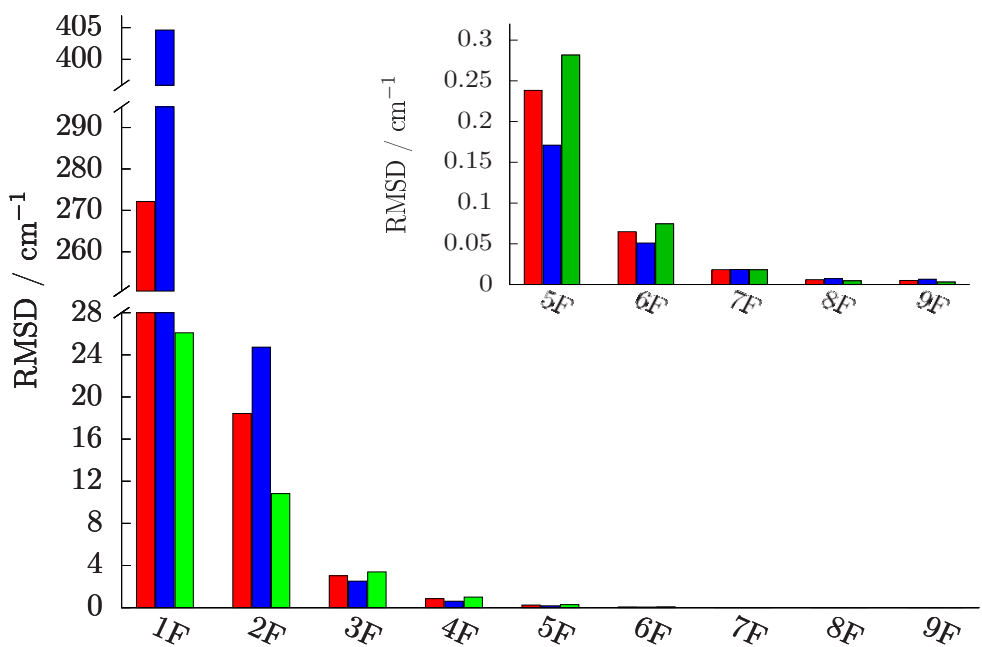}  
    \caption{RMSDs of fundamental excitations in icosadecaene
    obtained with DIF-ADGA and supersystem ADGA. 
    The molecule is fragmented into $\ce{-CH=CH-}$ units. 
    The large plot presents values calculated using 1M$l$F PESs with the truncation order $l$=1--9, while smaller
    plot shows enlarged area of $l$=5--9. Red, blue, and green colors present values 
    obtained for the entire set of modes (IC and INTRA) as well as for  
    IC and INTRA modes, respectively. All values are given in cm$^{-1}$.
    } \label{fig:conv_small}
\end{figure}

We have carried out DIF-ADGA calculations for the icosadecaene molecule using different sizes of molecular fragments 
and different truncation orders $l$ from one to $N-1$, where $N$ is the number of fragments. Note that 
the neighbor coupling approximation was employed in all PES calculations.
The resulting 1M$l$F PESs were used in subsequent VSCF calculations of fundamental excitation energies and compared to
excitation energies obtained based on the non-fragmented ADGA PES. RMSD values from supemolecular results are shown in 
Fig.~\ref{fig:conv_small}. 
The icosadecaene molecule was chosen for these computations as it features double bond 
conjugation and for that reason was expected to be very unfavorable in fragmentation schemes.
We anticipated that high truncation orders $l$ would be necessary to converge the resulting PES. 
However, as can be seen from Fig.~\ref{fig:conv_small} the calculated 1M$l$F PESs converge rather quickly.
The 1M1F PES introduces a very large RMSD of about 272 cm$^{-1}$ for the total set of vibrational modes due to a complete neglect of interactions 
between fragments. However, at higher truncation orders the RMSD decreases considerably and
reaches about 18 cm$^{-1}$ and 3.0 cm$^{-1}$ for $l=2$ and 3, respectively. Accuracy of 0.8 cm$^{-1}$ 
is essentially achieved for the 1M4F PES, while only negligible differences below 0.25 cm$^{-1}$ from the supersystem ADGA 
PES are found for higher orders of truncation, i.e., $l>4$. 
At low fragmentation orders, the RMSD values calculated for IC modes are much larger than those 
for the total set of modes and for INTRA vibrations. At the lowest truncation level $l=1$, the RMSD of IC vibrations is about 405 cm$^{-1}$.
However, this also quickly decreases with $l$ and drops below 0.6 cm$^{-1}$ for the 1M4F PES. 
It is interesting to note that the RMSDs of IC vibrations are much larger than those for INTRA modes at low orders of truncation $l<3$ only, whereas
the opposite behavior is observed for $l \ge 3$. This can probably be explained by the fact that although IC modes can span the entire molecule, 
in most cases they have large contributions at only 3--4 different fragments at once. 
Therefore, 
including FCs of order three leads to the convergence of many
IC cut-potentials and considerably decreases their RMSDs compared to those for INTRA modes.

\begin{figure}[h!]  
\centering 
\includegraphics[width=0.80\linewidth]{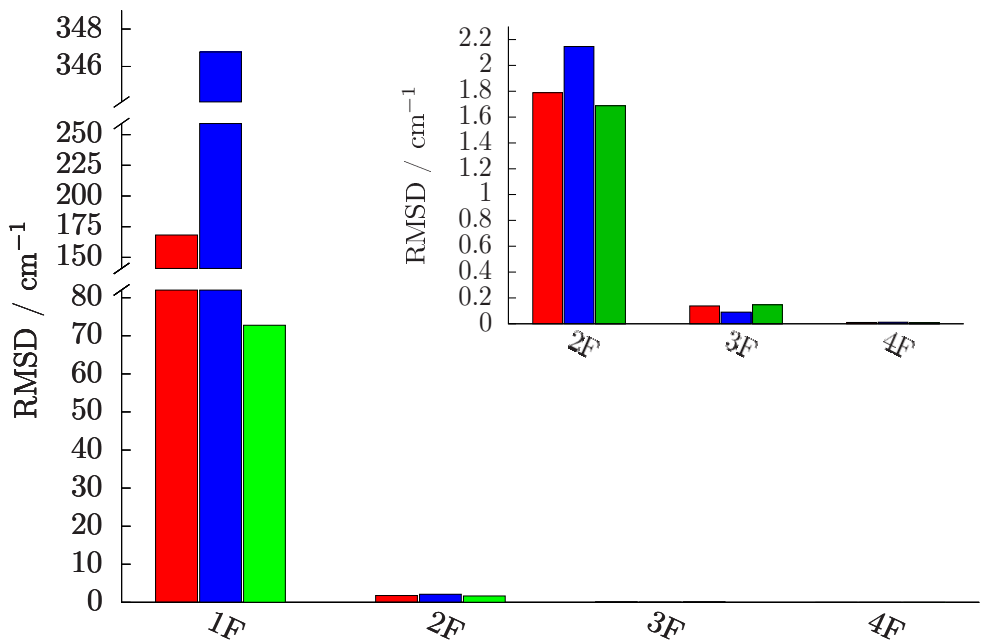}  
    \caption{RMSDs of fundamental excitations in icosadecaene
    obtained with DIF-ADGA and supersystem ADGA. 
    The molecule is fragmented into 
    $\ce{-(CH=CH)_2-}$
     units. 
    Large plot presents values calculated using 1M$l$F PESs with the truncation order $l$=1--4, while smaller
    plot shows enlarged area of $l$=2--4. Red, blue, and green colors present values 
    obtained for the entire set of modes (IC and INTRA) as well as for  
    IC and INTRA modes, respectively. All values are given in cm$^{-1}$.
    } \label{fig:conv_large}
\end{figure}

To further explore the convergence of the double incremental expansion, we carried out calculations for 
the icosadecaene molecule using twice as large molecular fragments
$\ce{-(CH=CH)_2-}$
than previously.
Note that these calculations were conducted using a different set of FALCON coordinates optimized for
these new fragments.
Therefore, the 1M$4$F PES with small molecular fragments 
$\ce{-CH=CH-}$ is not equivalent to the 1M$2$F PESs employing fragmentation of the total molecule 
into $\ce{-(CH=CH)_2-}$ units.
The results of these computations are shown in Fig.~\ref{fig:conv_large}. 
As expected, a very large RMSD value of about 168 cm$^{-1}$ is obtained for the total set of vibrational modes 
in the case of $l=1$. Higher truncation levels
exhibit decreasing RMSDs with errors of around 1.8 cm$^{-1}$ and 0.1 cm$^{-1}$ for $l=2$ and 3, respectively.
Therefore, the use of larger molecular fragments, indeed, results in a much faster convergence of the double 
incremental expansion in terms of the truncation order $l$. It is, however, interesting to note that 
the error of the 1M4F PES employing 
$\ce{-CH=CH-}$ fragments is equal to 0.8 cm$^{-1}$ 
(see Fig.~\ref{fig:conv_small}) and that this is about half of that obtained for the 1M2F 
PES with larger molecular fragments. These different RMSDs can probably be explained by the fact 
that in the latter case FCs 
$\ce{H-(CH=CH)_2-H}$ and $\ce{H-(CH=CH)_4-H}$ are calculated, whereas  
 in the former case, the FCs of order $l-1$ are, hence of the type 
 $\ce{H-(CH=CH)_3-H}$, c.f., Eq.~(\ref{eq:p_coeffs}). 
Accordingly, the use of small molecular fragments  
$\ce{-CH=CH-}$ and the truncation order $l=4$ in calculations of the 1M4F PESs results in computations 
of FCs composed of some larger molecular units than for the 1M2F PES with larger initial subsystems. 
 This leads to a better description of the supersystem PES 
and the smaller RMSDs obtained using the 1M4F PES.
Similar to the previous calculations, as depicted in Fig.~\ref{fig:conv_small}, the RMSDs for IC modes are larger than 
those for INTRA vibrations for small truncation orders $l<3$, whereas this situation is opposite for $l \ge 3$.

As shown above, the use of the truncation order $3 \le l \le 4$ in the many-body expansion is 
sufficient to obtain accuracy within 1~cm$^{-1}$ 
for the resulting fundamental excitation energies of the icosadecaene molecule.
The fast convergence of icosadecaene 1M$l$F PESs with respect to $l$ 
is rather surprising, when taking into account that the 
fragments are of a relatively modest size and only include 6--10 atoms, while having strong inter-fragment interactions.  
It can, therefore, be assumed that even small truncation orders $l$ might be  sufficient
for more favourable fragmentation cases. 
This, in fact, was already observed for tetraphenyl and hexaphenyl 
in Sec.~\ref{sec:validation}.

\section{Conclusions and Outlook \label{sec:concl}}

In this work, we presented the DIF-ADGA for PES construction. This methodology combines 
the cost-effective double incremental expansion of PESs with an automatic grid construction procedure, 
which allows for robust and fully-automated PES construction with a considerably reduced computational 
cost compared to a static grid approach. We described the  
DIF-ADGA implementation and thoroughly tested the method.
To that end, we compared the DIF-ADGA to 
other fragmented and non-fragmented computational methods by calculating fundamental 
excitation energies for a small set of chain-like molecular systems such as dicyclopropyl ketone, 
tetraphenyl, and hexaphenyl. 
For all three molecules under investigation, noticeable computational savings compared to a static 
grid approach were observed, while maintaining the same level of accuracy of the resulting PESs.
The most impressive results were obtained for the tetraphenyl and hexaphenyl molecules, 
where less than half SPs were calculated with the DIF-ADGA compared to DIF-Static. 
We also showed that the use of the neighbor coupling approximation within the DIF-ADGA 
results in only two to three times larger number of SPs than in
the standard non-fragmented ADGA. This is especially remarkable, when taking into account that
the DIF-ADGA SPs are considerably less expensive than those calculated for the non-fragmented molecular system.
For the hexaphenyl and tetraphenyl molecules,
estimates of the total computational time for PES construction were shown to be reduced by factors of 
2--3 and 7--41, when using the DIF-ADGA compared to the DIF-Static approach and supersystem static grid approach, 
respectively.

To further investigate the double incremental expansion of PESs and their convergence
at high orders of 
truncation in
the many-body expansion, we also carried out test calculations
for the long chain-like molecule of icosadecaene using differently sized 
fragments and truncation orders. 
Although this molecule features double bond conjugation, 
surprisingly fast convergence 
in the double incremental expansion
with the truncation order was observed.
Thus, deviations below 1 cm$^{-1}$ were obtained already at truncation of forth 
order in the expansion (1M4F PES), 
while using rather small fragments $\ce{-CH=CH-}$.
For larger fragments $\ce{-(CH=CH)_2-}$, similar accuracy was achieved
already for the third order expansion (1M3F PES). 
This proves that even for very unfavorable cases for fragmentation 
a fast convergence of the expansion can be obtained with the DIF-ADGA.

Therefore, we consider the DIF-ADGA to be a powerful tool for fully automatized and cost-efficient PES 
construction suitable for highly accurate vibrational spectra calculations.
This encourages 
us to further develop this methodology and 
to apply it in the context of
larger and more diverse 
molecular systems. This, however, requires a number of methodological issues to be solved. 
A further reduction of the computational cost can be achieved
by using either smart screening algorithms, which omit mode-coupling cut-potentials with
minor contributions to the total PES, 
or by implementing machine learning algorithms to predict
higher order surfaces.
The former approach is currently under development, whereas the latter strategy 
was recently probed for a set of small-sized molecules in the context of the standard 
ADGA and led to very high accuracy of the resulting PESs
for a much smaller computational cost~\cite{schmitz2019}.
Even larger computational savings can be reached by introducing the 
so-called double incremental expansion in FALCON coordinates
with auxiliary coordinate transformation (DIFACT). This theoretical method employs an approximate 
procedure for representing IC vibrational modes in terms of 
auxilliary coordiates that are purely local at molecular fragments and 
has been successfully applied in Ref.~\cite{koenig2016}. 
The use of DIFACT resulted in a linear-scaling computational cost with respect to the system size.
However, the additional approximation led to small errors in 
the PES compared to DIF. Integration of DIFACT and the ADGA would allow for PES calculations of much larger molecular systems
than presently accessible.
All above-mentioned techniques are currently under our active development and their combination with the ADGA will soon be available 
in {\sc MidasCpp}. 

\newpage
\clearpage

\section*{Acknowledgments}
We thank Diana Madsen for providing us with the tetraphenyl and hexaphenyl FALCON coordinates and
Leila Dzabbarova for exploratory computations on convergence of double incremental expansions.
D.G.A. acknowledges funding from the European Union's Horizon 2020 research and innovation
programme under the Marie Sk\l{}odowska--Curie grant agreement no.\ 835776.
C.K.\ acknowledges support from the Deutsche Forschungsgemeinschaft (DFG) through the Emmy Noether Young Group Leader Programme (project KO 5423/1-1). 
O.C.\ acknowledges support from the Danish Council for Independent Research through a Sapere Aude III grant (DFF-4002-00015). 

\newpage
\clearpage

\bibliographystyle{jcp}

\begin{thebibliography}{10}

\bibitem{galli1996}
G.~Galli.
\newblock {Linear scaling methods for electronic structure calculations and
  quantum molecular dynamics simulations}.
\newblock {\em Curr. Opin. Solid St. M.}, {\bf 1} (1996) 864--874.

\bibitem{goedecker1999}
S.~Goedecker.
\newblock {Linear scaling electronic structure methods}.
\newblock {\em Rev. Mod. Phys.}, {\bf 71} (1999) 1085--1123.

\bibitem{goedecker2003}
S.~Goedecker, G.E. Scuseria.
\newblock {Linear Scaling Electronic Structure Methods in Chemistry and
  Physics}.
\newblock {\em Comput. Sci. Eng.}, {\bf 5} (2003) 14--21.

\bibitem{zalesny2011}
R.~Zalesny, M.~G. Papadopoulos, P.~G. Mezey, J.~Leszczynski.
\newblock {\em Linear-Scaling Techniques in Computational Chemistry and
  Physics: Methods and Applications}.
\newblock Springer Netherlands, 2011.

\bibitem{sparta2009}
M.~Sparta, D.~Toffoli, O.~Christiansen.
\newblock {An adaptive density-guided approach for the generation of potential
  energy surfaces of polyatomic molecules}.
\newblock {\em Theor. Chem. Acc.}, {\bf 123} (2009) 413--429.

\bibitem{richter2012}
F.~Richter, P.~Carbonniere, A.~Dargelos, C.~Pouchan.
\newblock {An adaptive potential energy surface generation method using
  curvilinear valence coordinates}.
\newblock {\em J. Chem. Phys.}, {\bf 136} (2012) 224105.

\bibitem{strobusch2014}
D.~Strobusch, Ch. Scheurer.
\newblock {Adaptive sparse grid expansions of the vibrational Hamiltonian}.
\newblock {\em J. Chem. Phys.}, {\bf 140} (2014) 074111.

\bibitem{jung1996}
J.~O. Jung, R.~B. Gerber.
\newblock {Vibrational wave functions and spectroscopy of (H$_2$O)$_n$,
  $n$=2,3,4,5: Vibrational self‐consistent field with correlation
  corrections}.
\newblock {\em J. Chem. Phys.}, {\bf 105} (1996) 10332.

\bibitem{carter1997}
S.~Carter, S.~J. Culik, J.~M. Bowman.
\newblock {Vibrational self-consistent field method for many-mode systems: A
  new approach and application to the vibrations of CO adsorbed on Cu(100)}.
\newblock {\em J. Chem. Phys.}, {\bf 107} (1997) 10458.

\bibitem{bowman2003}
J.~M. Bowman, S.~Carter, X.~C. Huang.
\newblock {MULTIMODE: A code to calculate rovibrational energies of polyatomic
  molecules}.
\newblock {\em Int. Rev. Phys. Chem.}, {\bf 22} (2003) 533--549.

\bibitem{rauhut2004}
G.~Rauhut.
\newblock {Efficient calculation of potential energy surfaces for the
  generation of vibrational wave functions}.
\newblock {\em J. Chem. Phys.}, {\bf 121} (2004) 9313--9322.

\bibitem{kongsted2006}
J.~Kongsted, O.~Christiansen.
\newblock {Automatic generation of force fields and property surfaces for use
  in variational vibrational calculations of anharmonic vibrational energies
  and zero-point vibrational averaged properties}.
\newblock {\em J. Chem. Phys.}, {\bf 125} (2006) 124108.

\bibitem{rabitz1999}
H.~Rabitz, \"O.~F. Ali\c{s}.
\newblock {General foundations of high‐dimensional model representations}.
\newblock {\em J. Math. Chem.}, {\bf 25} (1999) 197--233.

\bibitem{meyer2012}
H.-D. Meyer.
\newblock {Studying molecular quantum dynamics with the multiconfiguration
  time‐dependent Hartree method}.
\newblock {\em Wiley Interdiscip. Rev.: Comput. Mol. Sci.}, {\bf 2} (2012)
  351--374.

\bibitem{benoit2004}
D.~M. Benoit.
\newblock {Fast vibrational self-consistent field calculations through a
  reduced mode–mode coupling scheme}.
\newblock {\em J. Chem. Phys.}, {\bf 120} (2004) 562--573.

\bibitem{benoit2006}
D.~M. Benoit.
\newblock {Efficient correlation-corrected vibrational self-consistent field
  computation of OH-stretch frequencies using a low-scaling algorithm}.
\newblock {\em J. Chem. Phys.}, {\bf 125} (2006) 244110.

\bibitem{pele2008}
L.~Pele, R.~B. Gerber.
\newblock {On the number of significant mode--mode anharmonic couplings in
  vibrational calculations: Correlation-corrected vibrational self-consistent
  field treatment of di-, tri-, and tetrapeptides}.
\newblock {\em J. Chem. Phys.}, {\bf 128} (2008) 165105.

\bibitem{benoit2008}
D.~M. Benoit.
\newblock {Fast vibrational calculation of anharmonic OH-stretch frequencies
  for two low-energy noradrenaline conformers}.
\newblock {\em J. Chem. Phys.}, {\bf 129} (2008) 234304.

\bibitem{seidler2009}
P.~Seidler, T.~Kaga, K.~Yagi, O.~Christiansen, K.~Hirao.
\newblock {On the coupling strength in potential energy surfaces for
  vibrational calculations}.
\newblock {\em Chem. Phys. Lett.}, {\bf 483} (2009) 138--142.

\bibitem{cheng2014}
X.~Cheng, R.~P. Steele.
\newblock {Efficient anharmonic vibrational spectroscopy for large molecules
  using local-mode coordinates}.
\newblock {\em J. Chem. Phys.}, {\bf 141} (2014) 104105.

\bibitem{yagi2007}
K.~Yagi, S.~Hirata, K.~Hirao.
\newblock {Multiresolution potential energy surfaces for vibrational state
  calculations}.
\newblock {\em Theor. Chem. Acc.}, {\bf 118} (2007) 681--691.

\bibitem{rauhut2008}
G.~Rauhut, T.~Hrenar.
\newblock {A combined variational and perturbational study on the vibrational
  spectrum of P$_2$F$_4$}.
\newblock {\em Chem. Phys.}, {\bf 346} (2008) 160--166.

\bibitem{Rauhut2009}
G.~Rauhut, B.~Hartke.
\newblock {Modeling of high-order many-mode terms in the expansion of
  multidimensional potential energy surfaces: Application to vibrational
  spectra}.
\newblock {\em J. Chem. Phys.}, {\bf 131}(1) (2009) 014108.

\bibitem{sparta2009_2}
M.~Sparta, I.-M. H{\o}yvik, D.~Toffoli, O.~Christiansen.
\newblock {Potential Energy Surfaces for Vibrational Structure Calculations
  from a Multiresolution Adaptive Density-Guided Approach: Implementation and
  Test Calculations}.
\newblock {\em J. Phys. Chem. A}, {\bf 113} (2009) 8712--8723.

\bibitem{sparta2010}
M.~Sparta, M.~B. Hansen, E.~Matito, D.~Toffoli, O.~Christiansen.
\newblock {Using Electronic Energy Derivative Information in Automated
  Potential Energy Surface Construction for Vibrational Calculations}.
\newblock {\em J. Chem. Theory Comput.}, {\bf 6} (2010) 3162--3175.

\bibitem{meier2013}
P.~Meier, G.~Bellchambers, J.~Klepp, F.~R. Manby, G.~Rauhut.
\newblock {Modeling of high-order terms in potential energy surface expansions
  using the reference-geometry Harris--Foulkes method}.
\newblock {\em Phys. Chem. Chem. Phys.}, {\bf 15} (2013) 10233--10240.

\bibitem{schmitz2019}
G.~Schmitz, D.~G Artiukhin, O.~Christiansen.
\newblock {Approximate high mode coupling potentials using Gaussian process
  regression and adaptive density guided sampling}.
\newblock {\em J. Chem. Phys.}, {\bf 150} (2019) 131102.

\bibitem{mackeprang2015}
K.~Mackeprang, V.~H\"anninen, L.~Halonen, H.~G. Kjaergaard.
\newblock {The effect of large amplitude motions on the vibrational intensities
  in hydrogen bonded complexes}.
\newblock {\em J. Chem. Phys.}, {\bf 142} (2015) 094304.

\bibitem{yagi2019}
K.~Yagi, K.~Yamada, C.~Kobayashi, Y.~Sugita.
\newblock {Anharmonic Vibrational Analysis of Biomolecules and Solvated
  Molecules Using Hybrid QM/MM Computations}.
\newblock {\em J. Chem. Theory Comput.}, {\bf 15} (2019) 1924--1938.

\bibitem{klinting2018}
E.~L. Klinting, B.~Thomsen, I.~H. Godtliebsen, O.~Christiansen.
\newblock {Employing general fit-bases for construction of potential energy
  surfaces with an adaptive density-guided approach}.
\newblock {\em J. Chem. Phys.}, {\bf 148} (2018) 064113.

\bibitem{sparta_mult2009}
M.~Sparta, I.-M. H{\o}yvik, D.~Toffoli, O.~Christiansen.
\newblock {Potential Energy Surfaces for Vibrational Structure Calculations
  from a Multiresolution Adaptive Density-Guided Approach: Implementation and
  Test Calculations}.
\newblock {\em J. Phys. Chem. A}, {\bf 113} (2009) 8712--8723.

\bibitem{koenig2016}
C.~K\"onig, O.~Christiansen.
\newblock {Linear-scaling generation of potential energy surfaces using a
  double incremental expansion}.
\newblock {\em J. Chem. Phys.}, {\bf 145} (2016) 064105.

\bibitem{koenig2016_falcon}
C.~K\"onig, M.~B. Hansen, I.~H. Godtliebsen, O.~Christiansen.
\newblock {FALCON: A method for flexible adaptation of local coordinates of
  nuclei}.
\newblock {\em J. Chem. Phys.}, {\bf 144} (2016) 074108.

\bibitem{madsen2018}
D.~Madsen, O.~Christiansen, C.~K\"onig.
\newblock {Anharmonic vibrational spectra from double incremental potential
  energy and dipole surfaces}.
\newblock {\em Phys. Chem. Chem. Phys.}, {\bf 20} (2018) 3445--3456.

\bibitem{toffoli2011}
D.~Toffoli, M.~Sparta, O.~Christiansen.
\newblock {Accurate Multimode Vibrational Calculations Using A B-spline Basis:
  Theory, Tests and Application to Dioxirane and Diazirinone}.
\newblock {\em Mol. Phys.}, {\bf 109} (2011) 673--685.

\bibitem{orca}
F.~Neese.
\newblock {The ORCA program system}.
\newblock {\em Comput. Mol. Sci.}, {\bf 2} (2012) 73--78.

\bibitem{sure2013}
R.~Sure, S.~Grimme.
\newblock {Corrected small basis set Hartree‐-Fock method for large systems}.
\newblock {\em J. Comput. Chem.}, {\bf 34} (2013) 1672--1685.

\bibitem{midascpp}
O.~Christiansen, D.~G. Artiukhin, I.~H. Godtliebsen, E.~M. Gras,
  W.~Gy{\H{o}}rffy, M.~B. Hansen, M.~B. Hansen, E.~L. Klinting, J.~Kongsted,
  C.~K{\"o}nig, D.~Madsen, N.~K. Madsen, K.~Monrad, G.~Schmitz, P.~Seidler,
  K.~Sneskov, M.~Sparta, B.~Thomsen, D.~Toffoli, A.~Zoccante.
\newblock {MidasCpp}, version 2019.07.0.
\newblock https://midascpp.gitlab.io/.

\bibitem{toffoli2007}
D.~Toffoli, J.~Kongsted, O.~Christiansen.
\newblock {Automatic generation of potential energy and property surfaces of
  polyatomic molecules in normal coordinates}.
\newblock {\em J. Chem. Phys.}, {\bf 127} (2007) 204106.

\bibitem{bowman1978}
J.~M. Bowman.
\newblock {Self‐consistent field energies and wavefunctions for coupled
  oscillators}.
\newblock {\em J. Chem. Phys.}, {\bf 68} (1978) 608--610.

\bibitem{gerber1979}
R.~B. Gerber, M.~A. Ratner.
\newblock {A semiclassical self-consistent field (SC SCF) approximation for
  eigenvalues of coupled-vibration systems}.
\newblock {\em Chem. Phys. Lett.}, {\bf 68} (1979) 195--198.

\bibitem{christiansen2004}
O.~Christiansen.
\newblock {A second quantization formulation of multimode dynamics}.
\newblock {\em J. Chem. Phys.}, {\bf 120} (2004) 2140--2148.

\bibitem{hansen2010}
M.~B. Hansen, M.~Sparta, P.~Seidler, D.~Toffoli, O.~Christiansen.
\newblock {New Formulation and Implementation of Vibrational Self-Consistent
  Field Theory}.
\newblock {\em J. Chem. Theory Comput.}, {\bf 6} (2010) 235--248.

\bibitem{beck1988}
A.~D. Becke.
\newblock {Density-functional exchange-energy approximation with correct
  asymptotic behavior}.
\newblock {\em Phys. Rev. A}, {\bf 38}(6) (1988) 3098--3100.

\bibitem{perd1986}
J.~P. Perdew.
\newblock {Density-functional approximation for the correlation energy of the
  inhomogeneous electron gas}.
\newblock {\em Phys. Rev. B}, {\bf 33} (1986) 8822--8824.

\bibitem{schaefer1992}
A.~Sch\"afer, H.~Horn, R.~Ahlrichs.
\newblock {Fully optimized contracted Gaussian basis sets for atoms Li to Kr}.
\newblock {\em J. Chem. Phys.}, {\bf 97} (1992) 2571--2577.

\bibitem{schaefer1994}
A.~Sch\"afer, C.~Huber, R.~Ahlrichs.
\newblock {Fully optimized contracted Gaussian basis sets of triple zeta
  valence quality for atoms Li to Kr}.
\newblock {\em J. Chem. Phys.}, {\bf 100} (1994) 5829--5835.

\bibitem{eichkorn1995}
K.~Eichkorn, O.~Treutler, H.~\"Ohm, M.~H\"aser, R.~Ahlrichs.
\newblock {Auxiliary basis sets to approximate Coulomb potentials}.
\newblock {\em Chem. Phys. Letters}, {\bf 240} (1995) 283--290.

\bibitem{eichkorn1997}
K.~Eichkorn, F.~Weigend, O.~Treutler, R.~Ahlrichs.
\newblock {Auxiliary basis sets for main row atoms and transition metals and
  their use to approximate Coulomb potentials}.
\newblock {\em Theor. Chem. Acc.}, {\bf 97} (1997) 119--124.


\end{thebibliography}

\end{document}